\documentclass{article}
\usepackage{spconf, epsf,epsfig,bezier,amsmath,amssymb,array,color,subfig}
\usepackage{graphicx,amsfonts,latexsym}
\usepackage{epstopdf, times}
\usepackage{url}
\usepackage{hyperref}

\usepackage{fancyhdr}

\title{Information Decoding and SDR Implementation of DFRC Systems Without Training Signals}
\name{Daniel M. Wong$^\dag$, Batu K. Chalise$^\dag$, Justin Metcalf$^\ddag$, Moeness Amin$^{*}$
}
\address{$^\dag$Electical and Computer Engineering Department, New York Institute of Technology,\\
Old Westbury, NY 11568, emails: {\it \{dwong07, bchalise\}@nyit.edu}\\
$^\ddag$ Advanced Radar Research Center, University of Oklahoma \\
  Norman, OK 73019, email: {\it jmetcalf@ou.edu}\\
  $^{*}$ Center of Advanced Communications, Villanova University \\
  Villanova, PA 19087, email: {\it moeness.amin@villanova.edu}
}

\begin{document}
\maketitle
\thispagestyle{empty}

 
 
 \begin{abstract}
  Recent performance analysis of dual-function radar communications (DFRC) systems, which embed information using phase shift keying (PSK) into multiple-input multiple-output (MIMO) frequency hopping (FH) radar pulses, shows promising results for addressing spectrum sharing issues between radar and communications. However, the problem of decoding information at the communication receiver remains challenging, since the DFRC transmitter is typically assumed to transmit only information embedded radar waveforms and not the training sequence. We propose a novel method for decoding information at the communication receiver without using training data, which is implemented using a software-defined radio (SDR). The performance of the SDR implementation is examined in terms of  bit error rate (BER) as a function of signal-to-noise ratio (SNR) for  differential binary and quadrature phase shift keying modulation schemes and compared with the BER versus SNR obtained with numerical simulations.   
  
\end{abstract}
 

 \vspace*{-0.3cm}

\section{Introduction}
\label{secno1}
The {\it co-existence} and {\it co-design} paradigms for joint radar and communications have recently gained significant research interest due to their potential to mitigate spectrum congestion. The co-existence approach enables frequency spectrum sharing between separately developed radar and communications systems by limiting interference caused by one system to another \cite{MartoneAES18}-\cite{MartoneRadarConf16}. The co-design approach focuses on using transmitters and/or receivers on the same platform to function as both radar and communications systems \cite{ChaliseHimedSPL17}-\cite{ChaliseAminDSP18}. 

A typical objective in the co-design approach is to develop a joint transmitter that embeds information intended for communications receivers into the radar waveforms intended for radar receivers \cite{AminSPM19}-\cite{AminIndu19}. For this purpose,  intentional modulation is employed, where the original radar waveform, having a wider bandwidth with a faster Nyquist sampling rate, is remodulated by the desired communication waveform, having a much narrower bandwidth with a much slower Nyquist sampling rate \cite{HassanienRadarConf19}, \cite{WuCommMag16}{\footnote{Different from remodulation concept, some systems use radar waveforms to mask information signals for providing covertness in communications \cite{Tedesso}.}}. Towards this end, intrapulse modulation methods, namely, eigenvectors-based method, weighted-combining, dominant-projection, and radar waveform sidelobe modulation have been proposed in \cite{BluntAES10} and \cite{AhmadIET16}. Although other modulation schemes, such as binary phase shift keying (BPSK), quadrature phase shift keying (QPSK), and multilevel M-ary PSK are also proposed to remodulate the radar waveforms,  these methods impede the radar functionality. Their adverse effects lie in increased spectral leakage and inferior ambiguity function \cite{HassanienRadarConf19}. Radar waveforms, such as linear frequency modulation (LFM), nonlinear frequency modulation (NLFM), phase-coded waveforms, etc., are generally designed in such a way to achieve the desired peak-to-sidelobe ratio in pulse compression, as well as the desired spectral sidebands distribution to satisfy the requirements on radar spectrum. To remedy these  effects, a reduced phase modulation scheme was proposed in \cite{SahinRadarConf17}-\cite{SahinRadarConf18}.

 Despite the progress made, the aforementioned DFRC systems (for example see \cite{HassanienRadarConf19} and references therein) assume that the communications end users can decode their information utilizing channel state information (CSI) of the channels between the DFRC transmitters and communications users. However, to estimate this CSI at the users, the DFRC transmitter should transmit training signals, which requires additional transmit power, bandwidth and/or reduces data rate. Moreover, it is not clear how these training signals can be embedded into radar waveforms without affecting target detection and tracking capability of the radar. As such, it is beneficial to develop schemes that can decode information without requiring CSI estimation.      
 
In this paper, we propose a novel method of decoding information at the communication receiver without the need to estimate the CSI.  As such, no transmitted training signal is required by the DFRC platform. The information is embedded into radar waveforms employing differential BPSK (D-BPSK) and differential QPSK (D-QPSK) modulation schemes. The proposed method assumes that the DFRC transmitter is equipped with multiple antennas, whereas the communication receiver has a single antenna. The DFRC transmitter generates frequency hopping (FH) waveforms. A software-defined radio (SDR) implementation, with USRP-2901 and LabVIEW software, is proposed to implement the decoding method for a special case of DFRC transmitter with a single antenna{\footnote{To the best of our knowledge, this is the first work  that deals with the SDR implementation of information decoding in the DFRC system. The closest paper to ours is \cite{RayIIT}, which proposes modulation of frequency modulated continuous wave (FMCW) radar signal with double side-band suppressed carrier method and a GNU-radio based simulator block diagram. However, the hardware implementation is not proposed, and therefore, a SDR implementation is identified only as a future work in \cite{RayIIT}.}}. The performance of the proposed SDR is demonstrated in terms of bit error rate (BER) versus signal-to-noise ratio (SNR) and compared with that obtained from MATLAB-based Monte Carlo simulations.

The remainder of this paper is organized as follows. The system model of the FH-MIMO  DFRC system is described in Section \ref{secno2}. The proposed D-BPSK method to recover differential phase information without requiring CSI estimation is presented in \ref{secno3}. The SDR implementation and performance results are described in Sections  \ref{secno4} and  \ref{secno5}, respectively. Finally, the conclusions are drawn in Section  \ref{secno6}.

   
\vspace*{-0.3cm}
\section{System Model}
\label{secno2}
We consider a DFRC system consisting of co-located transmit and receive antennas. Let $M$  be the number of uniform-linearly spaced omnidirectional co-located transmit and antennas. For the $m$-th transmit antenna, the transmitted FH waveform can be expressed as
\vspace*{-0.2cm}
\begin{eqnarray}
\label{eqn1}
\phi_m(t)=\sum_{q=1}^{Q}e^{j2\pi c_{m,q} \Delta_f t} u(t-q\Delta_t),
\end{eqnarray}
where $c_{m,q}$ is the FH coefficient with $m=1, \cdots, M$ and $q=1, \cdots, Q$. Here, $Q$ is the number of sub-pulses within a FH waveform (pulse), $\Delta_f$ is the frequency step, $\Delta_t$ is the sub-pulse duration, and
\begin{eqnarray}
\label{eqn2}
u(t)=\biggl\{\begin{array}{cc} 1, & 0\leq t\leq \Delta_t \\ 0, & otherwise \\ \end{array}
\end{eqnarray}
is the rectangular pulse of duration $\Delta_t$. The FH coefficients are assumed to satisfy the following condition:
\begin{eqnarray}
\label{eqn3}
c_{m,q}\neq c_{m', q}, \forall m'\neq m, q, 
\end{eqnarray}
which means that  the sub-pulse frequencies can be equal across different sub-pulses, but not across different transmit antennas.
 
The set of possible PSK values, $D_{PSK}$, is defined as $\{0,\frac{2\pi}{J}, \cdots, \frac{(J-1)2 \pi}{J}\}$ for a dictionary size $J$. For the $n$-th pulse (slow time index), the PSK modulated frequency-hopped (FH/PSK) waveform is given by
\begin{eqnarray}
\label{eqn4}
\psi_m(t,n)=\sum_{q=1}^{Q} e^{j\Omega_{m,q}^{(n)}}h_{m,q}(t) u(t-q\Delta_t-nT_0),
\end{eqnarray}
where $h_{m,q}(t) =e^{j2\pi c_{m,q} \Delta_f t}$  is the FH waveform associated with the $q$-th sub-pulse of the $m$-th antenna, $T_0$ is the pulse repetition interval (PRI), and $\Omega_{m,q}^{(n)} \in D_{PSK}$.  

Consider a single-antenna communication receiver located at $\theta_c$ with respect to the DFRC transmitter. We assume non-frequency selective Rayleigh fading channel  between the transmitter and the communication receiver, which is valid when bandwidth of the FH/PSK waveform is smaller than the coherence bandwidth of the channel, i.e., in the narrowband assumption. The signal received by the communication receiver can be expressed as 
\begin{eqnarray}
\label{eqn7}
r(t,n)=\alpha_{ch} {\bf a}^T(\theta_c) {\boldsymbol \psi}(t,n)+w(t,n),
\end{eqnarray}
where ${\boldsymbol \psi}(t,n)=[ \psi_1(t,n), \cdots,   \psi_M(t,n) ]^T$, $\alpha_{ch}$ is the channel coefficient, ${\bf a}(\theta_c)$ is the $M\times 1$ steering vector corresponding to $\theta_c$, and $w(t,n)$ is the zero-mean additive white Gaussian noise with variance $\sigma_w^2$. We also assume that within a pulse duration, the change in channel coefficient and spatial direction is negligible.
Assuming time and phase synchronization between the DFRC transmitter and the communication receiver, matched filtering of (\ref{eqn7}) with the FH sub-pulses yields 
\begin{eqnarray}
\label{eqn8}
y_{m,q}&=&\int_{\Delta_t}^{} r(t,n)h^{*}_{m,q}(t) u(t-q\Delta_t -nT_0) \ dt, \nonumber\\
&=&\alpha_{ch} e^{j( \Omega_{m,q}^{(n)} -2\pi d_m \sin(\theta_c))}+w_{m,q}(n),
\end{eqnarray}
 where $d_m$  is the displacement between the first and $m$-th transmit array elements measured in wavelength, $w_{m,q}(n)=\int_{0}^{\Delta_t} w(t,n) 
 e^{-j2\pi c_{m, q}\Delta_f t} u(t-\Delta_t-n T_0) \ dt, $ is zero-mean additive noise at the output of the $(m,q)$-th match filter with variance $\sigma_m^2$. Under high SNR assumption,  $\Omega_{m,q}^{(n)}$ can be estimated as  
 \begin{eqnarray}
 \label{eqn9}
 {\hat \Omega}_{m,q}^{(n)}=\angle(y_{m,q}(n))-\psi_{ch}+2\pi d_m \sin(\theta_c),    
 \end{eqnarray} 
 where $\angle(y_{m,q}(n))$ is the angle $y_{m,q}(n)$,  and $\psi_{ch}$ is the phase of the channel coefficient, $\alpha_{ch}$. 
 \vspace*{-0.3cm}
\section{Proposed Decoding}
\label{secno3}
The two variables $\psi_{ch}$ and ${\theta}_c$ are typically assumed known \cite{HassanienRadarConf19} or estimated using a training sequence. However,  the training sequence  consumes  resources and constitutes undesirable overhead on the primary radar which is not designed to assist a  communication receiver. Nonetheless, after carefully analyzing (\ref{eqn8}), we can form phase difference between matched filter outputs of two consecutive sub-pulses and eliminate the two unknown channel parameters. To demonstrate this method, note that, under high SNR assumption, for the $q$-th and $(q+1)$-th sub-pulses,  (\ref{eqn8}) can be expressed as
\vspace*{-0.2cm}
\begin{eqnarray}
\label{eqn8A}
y_{m,q}& \approx &\alpha_{ch} e^{j( \Omega_{m,q}^{(n)} -2\pi d_m \sin(\theta_c))},\\
y_{m,q+1}& \approx &\alpha_{ch} e^{j( \Omega_{m,q+1}^{(n)} -2\pi d_m \sin(\theta_c))}.
\end{eqnarray}
 Clearly, the ratio $\frac{y_{m,q+1}(n)}{y_{m,q}(n)}$ reduces to $e^{j( \Omega_{m,q+1}^{(n)}-\Omega_{m,q}^{(n)})}$. As such, the phase difference between two consecutive sub-pulses is
 \vspace*{-0.3cm}
\begin{eqnarray}
\label{eqn10}
\Omega_{m,q+1}^{(n)}-\Omega_{m,q}^{(n)}=-j\ln \left(\frac{y_{m,q+1}(n)}{y_{m,q}(n)}\right),
\end{eqnarray}
where $q=1, \cdots, Q-1$.  It is clear from (\ref{eqn10}) that the unknown channel parameters do not need to be estimated for recovering the differential phase. In essence,  information decoding will be possible if the dual-function transmitter employs differential PSK modulation. In such modulation, the current phase value is given by $ \Omega^{(n)}_{m,q+1}= \Omega^{(n)}_{m,q}+\Delta\Omega^{(n)}_{m,q}$, where $\Delta\Omega^{(n)}_{m,q}$ is the differential phase. Since we consider D-BPSK and D-QPSK constellations,  $\Delta\Omega^{(n)}_{m,q}$ takes values of $\frac{\pi}{2}$ and $\frac{3\pi}{2}$ for D-BPSK and $\frac{\pi}{4}$, $\frac{3\pi}{4}$, $\frac{5\pi}{4}$, and $\frac{7\pi}{4}$ for D-QPSK. 

\vspace*{-0.3cm}
\section{Proposed SDR Implementation}
\label{secno4}
\begin{figure}[htb!]
 \centering
 \includegraphics[width=7cm]{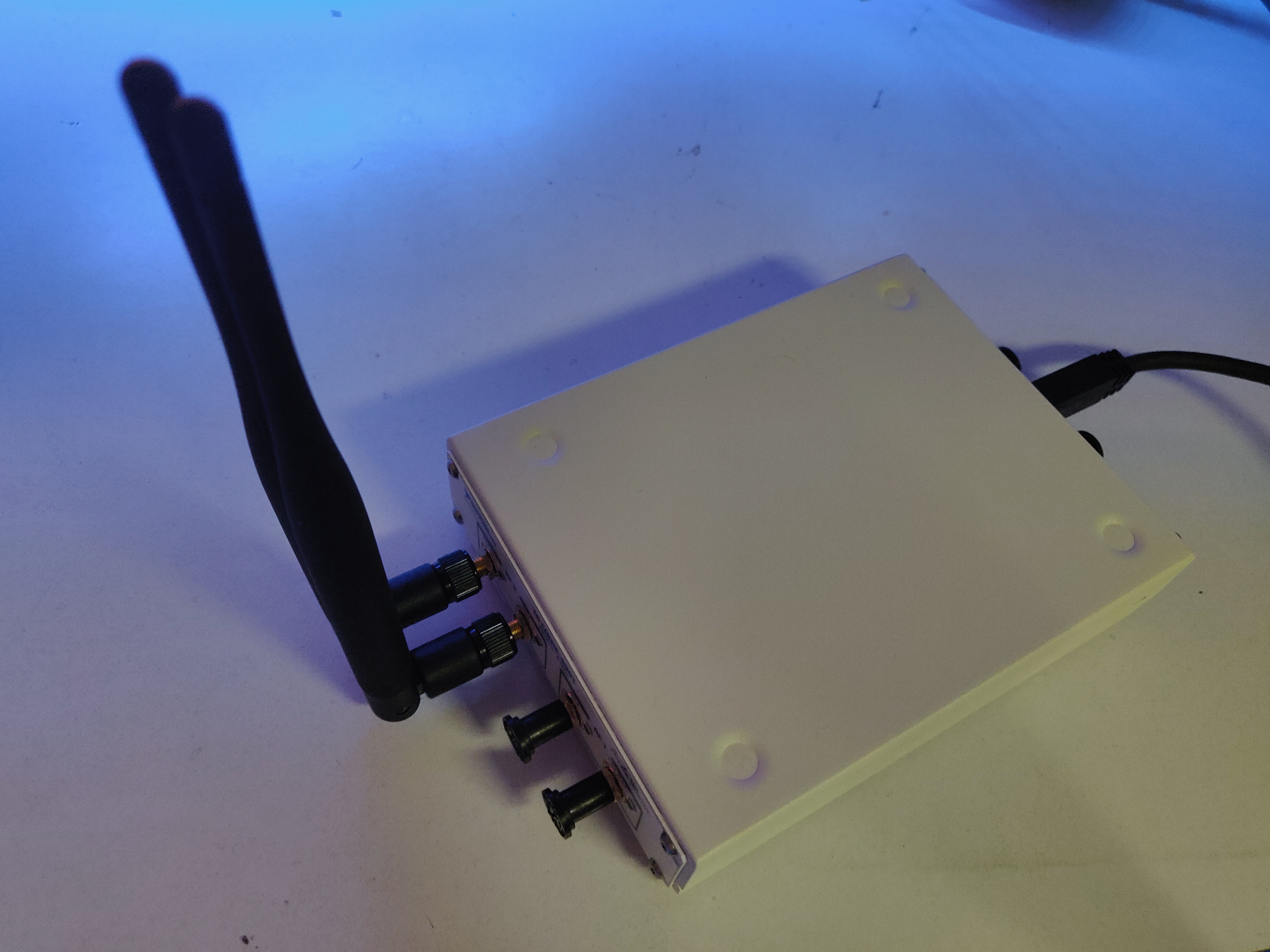}
  \caption{NI USRP-2901 using VERT2450 antennas}
  \label{fig-Air}
  \end{figure}

We implement the proposed information decoding using a SDR, which includes USRP-based hardware and LabVIEW-based software. Our hardware setup consists of a NI USRP-2901, which is connected to a computer over USB 3.0. We use antennas to test line of sight wireless transmissions. As shown in Figure \ref{fig-Air}, two VERT2450 dual-band antennas are used to transmit port TX1 and receive port RX2 on channel 0.
  
 The USRP parameters chosen to simulate the proposed model for both D-BPSK and D-QPSK modulations are as follows: RF channel 0, transmit port TX1, receive port RX2, 5 GHz carrier frequency, 30 MHz IQ rate, 20 MHz sampling rate, 1 dB receiver gain, 250 kHz FH interval, all FH code values of $10$, $10^3$ to $10^5$ signal pulses, and transmit gain of $-2.5$ dB to $7.5$ dB at an increment of 2.5 dB. To maintain the same energy per bit between D-BPSK and D-QPSK, we transmit twice as many samples in D-QPSK than in D-BPSK; with the same sampling frequency, D-BPSK uses a 1 us sub-pulse interval, whereas D-QPSK uses a 2 us sub-pulse interval.
 
\begin{figure}[t]
	\centering
	\subfloat[]{
		\includegraphics[width=0.8\columnwidth]{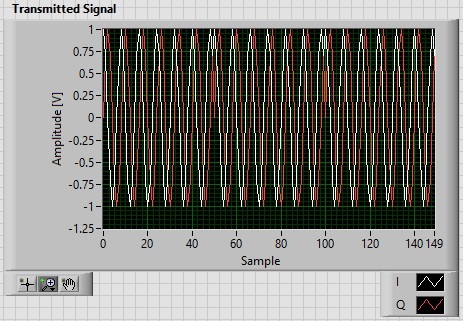}
		\label{fig:TxSignal}
		}
	\hskip2em
	\subfloat[]{
		\includegraphics[width=0.8\columnwidth]{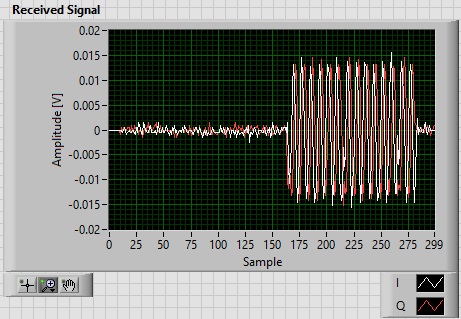}
		\label{fig:RxSignal}
		}		
	\caption{\small USRP example of 5 signal pulses (at transmitter gain 20 dB) of (a) transmitted signal (b) received signal.}
	\label{fig:TxRxSignal}
\end{figure}

The USRP-2901 is controlled using LabVIEW as the interface. Two .vi files, one for the transmitter and the other for the receiver, are used. In the transmitter .vi, we first generate the transmitted waveform (\ref{eqn4}) through a MATLAB script node. During the execution of the MATLAB script, randomly generated information bits are exported to a .mat file. After the script is finished executing, the transmitter port on the USRP-2901 is initialized. The transmitter is then configured to the appropriate gain, carrier frequency, and IQ rate. The FH/PSK waveform is then transmitted. In the receiver .vi, the receive port on the USRP is initialized and fetches the received signal. The received signal is exported as a .csv file. The synchronization between the transmitter and receiver does introduce a delay as seen in Figure \ref{fig:TxRxSignal}. 

A MATLAB script imports the transmitted bits and received waveforms to calculate the BER. The DC-bias removal and IQ-imbalance correction are performed on the received signals. By cross correlating the received signals with the FH waveform, the delay between the transmitter and receiver was determined. Matched filtering (\ref{eqn8}) is applied between the received signal and the FH waveform signal to estimate the embedded information bits. These estimated bits are compared with the actual bits and the BER is calculated.
\vspace*{-0.3cm}
\section{Results}
\label{secno5}
\begin{figure}[htb!]
 \centering
 \includegraphics[width=7cm]{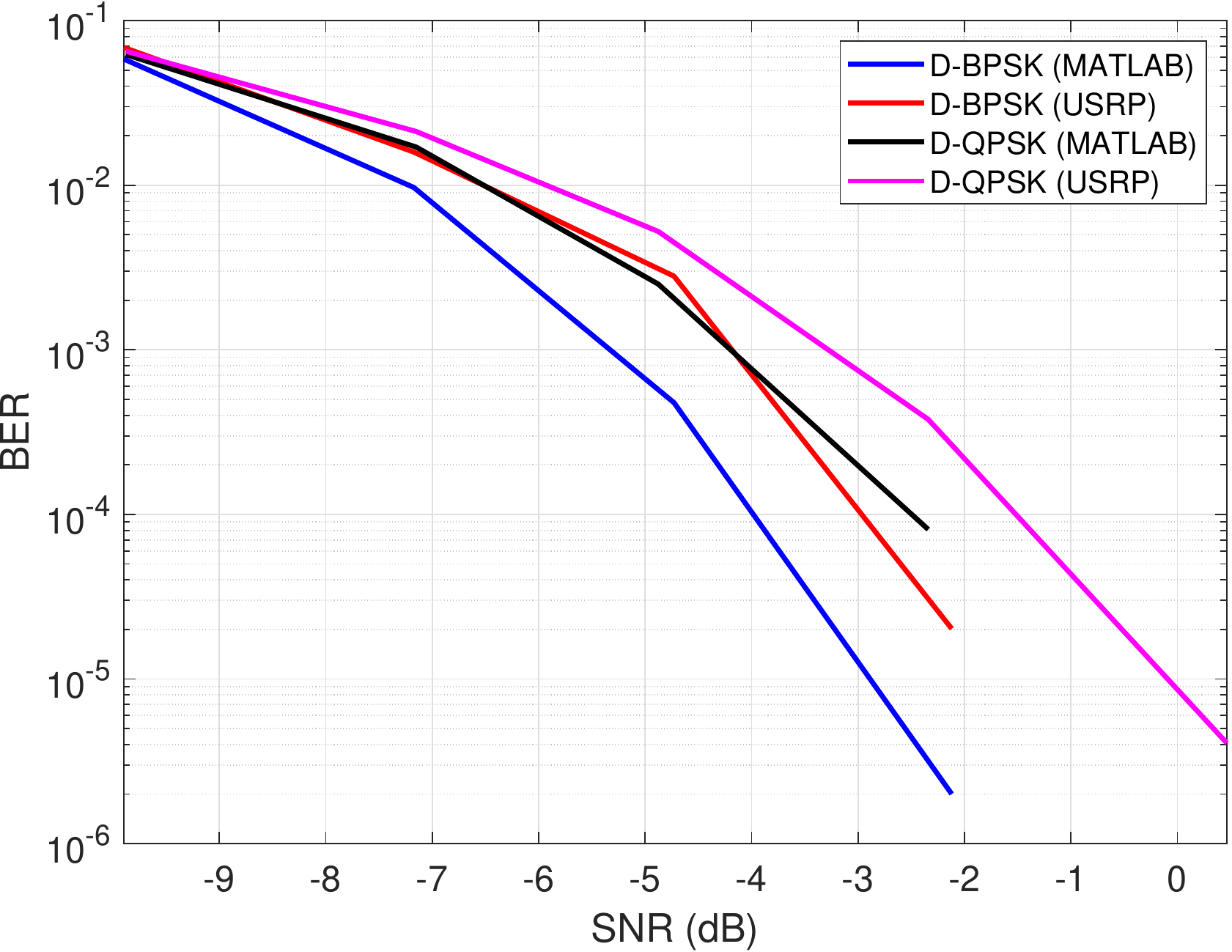}
  \caption{BER vs SNR for MATLAB and SDR}
  \label{fig-ResCable}
  \end{figure}

Our objective is to compare the performance of the proposed SDR implementation in terms of BER versus SNR with that obtained from ideal MATLAB-based Monte Carlo simulations. To this end,  the SNR corresponding to each USRP transmitter gain is calculated as
\begin{eqnarray}
\label{eqn11}
SNR=\frac{P_{noisy~signal}-P_{noise}}{P_{noise}}.
\end{eqnarray}
The power of noise, $P_{noise}$,  in (\ref{eqn11}) is calculated from the received signal when the USRP does not transmit signals, whereas the power of noisy signal,  $P_{noisy~signal}$, is calculated using the received signal when the USRP transmits the FH/PSK waveform. The same number of the samples of the received signals are used for calculating these powers. Figure \ref{fig-ResCable} compares the BER versus SNR curves obtained with MATLAB-based Monte Carlo runs and the proposed SDR-based experimental setup, when using D-BPSK and D-QPSK modulations. For the MATLAB simulations, the additive white Gaussian noise (AWGN) channel  is assumed, i.e., $\alpha_{ch}=1$. 

\begin{figure}[htb!]
 \centering
 \includegraphics[width=7cm]{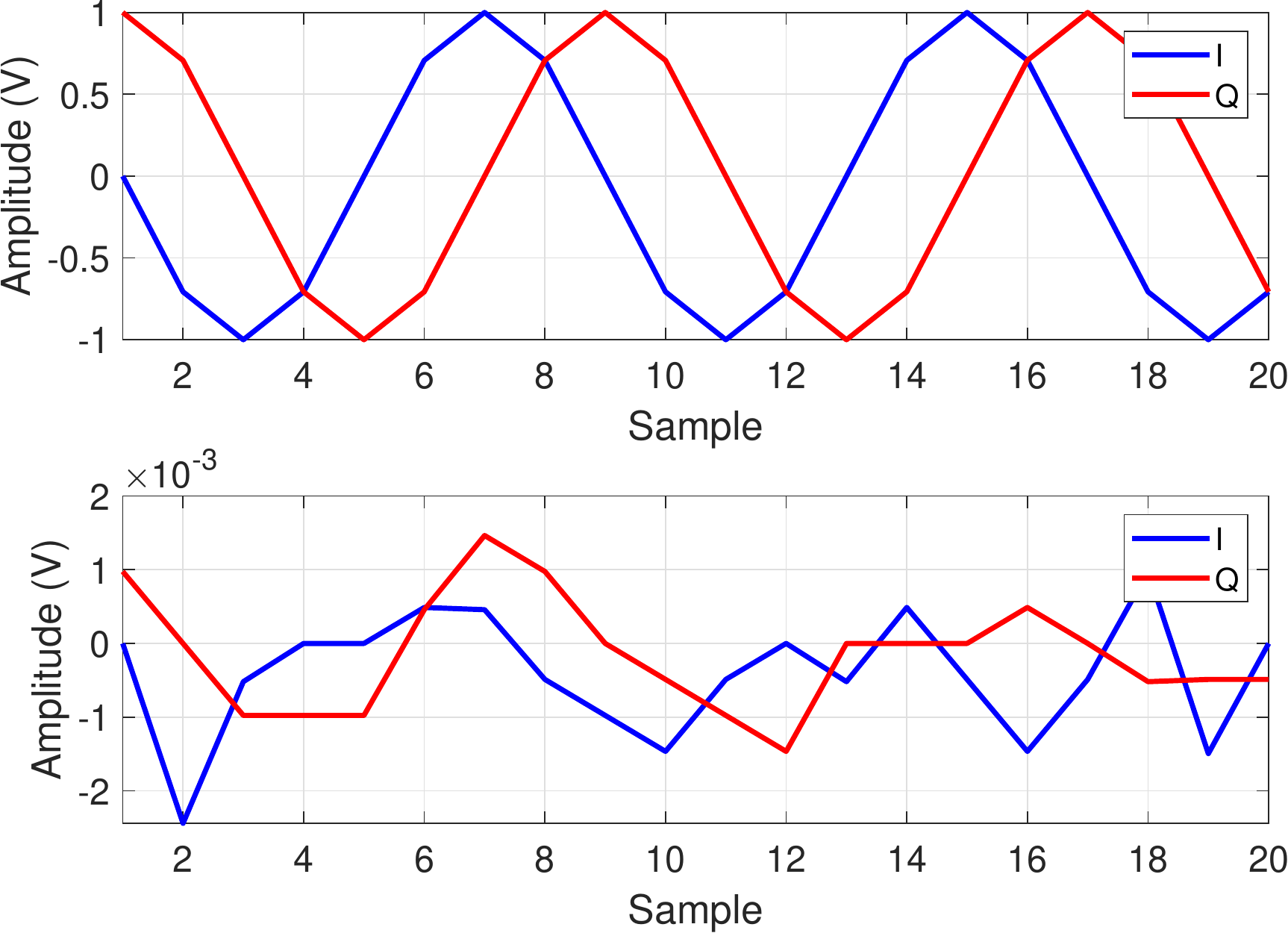}
  \caption{Discrepancy between transmitted (top) and received (bottom) signals for some runs}
  \label{fig-Distortion}
  \end{figure}

 It can be observed from this figure that the BER values obtained with the SDR implementation are comparable to those obtained with MATLAB simulations; the difference between MATLAB and SDR is within $1$ dB and the difference between D-BPSK and D-QPSK is within 3 dB. The slightly higher BER obtained from the SDR implementation can be attributed to discrepancies in the mapping between the transmit gain of the USRP and the calculated SNR. Moreover, due to limitations of LabVIEW and the host computer, only 20, 000 samples (i.e., 1,000 bits) are transmitted for each capture (run) of the received USRP signal. At transmitter gain values less than $-5$dB, we employed $1,000$ such runs to account for the transmission of $10^6$ bits in total. At higher gain values, we employed $10,000$ and $100,000$ runs. Since these runs cannot be replicated in identical conditions due to fluctuating noise and possible hardware impairments, the transmit gain to SNR mapping is subject to some errors. There is also signal distortion observed in some runs as seen in Figure \ref{fig-Distortion} with the in-phase and quadrature components of the transmitted and received signals. The received signal does not resemble the transmitted signal. This is also another reason for slightly higher BER value achieved with the proposed SDR implementation compared to ideal MATLAB simulations. 
 
 \vspace*{-0.3cm}
 \section{Conclusions}
  \label{secno6}
 We proposed a novel information decoding method for a frequency-hopped MIMO dual-function radar communications (DFRC) system in the absence of training signals, leveraging differential binary and quadrature phase shift keying modulation schemes. A software-defined radio (SDR), with USRP-2901 and LabVIEW software,  was implemented  for a DFRC transmitter with a single antenna. The performance of the proposed SDR, in terms of the bit error rate (BER) versus signal-to-noise ratio (SNR),  was found to be comparable with that obtained with the ideal MATLAB-based Monte Carlo simulations.

\end{document}